\documentclass[ALICE,manyauthors]{cernphprep}
\usepackage[comma,square,numbers,sort&compress]{natbib}
\usepackage{hyperref}
\usepackage{xspace}

\usepackage[utf8]{inputenc}
\usepackage{amssymb}
\usepackage{amsmath}
\usepackage{authblk}
\usepackage[T1]{fontenc}

\newcommand{\eq}[1]{\begin{align} #1 \end{align}}

\newcommand{\Brom}{\mathrm{B}} 
\newcommand{\Prom}{\mathrm{p}} 

\begin{document}
%

\newcommand{\pp}           {pp\xspace}
\newcommand{\ppbar}        {\mbox{$\mathrm {p\overline{p}}$}\xspace}
\newcommand{\XeXe}         {\mbox{Xe--Xe}\xspace}
\newcommand{\PbPb}         {\mbox{Pb--Pb}\xspace}
\newcommand{\pA}           {\mbox{pA}\xspace}
\newcommand{\pPb}          {\mbox{p--Pb}\xspace}
\newcommand{\AuAu}         {\mbox{Au--Au}\xspace}
\newcommand{\dAu}          {\mbox{d--Au}\xspace}

\newcommand{\s}            {\ensuremath{\sqrt{s}}\xspace}
\newcommand{\snn}          {\ensuremath{\sqrt{s_{\mathrm{NN}}}}\xspace}
\newcommand{\pt}           {\ensuremath{p_{\rm T}}\xspace}
\newcommand{\meanpt}       {$\langle p_{\mathrm{T}}\rangle$\xspace}
\newcommand{\ycms}         {\ensuremath{y_{\rm CMS}}\xspace}
\newcommand{\ylab}         {\ensuremath{y_{\rm lab}}\xspace}
\newcommand{\etarange}[1]  {\mbox{$\left | \eta \right |~<~#1$}}
\newcommand{\yrange}[1]    {\mbox{$\left | y \right |~<~#1$}}
\newcommand{\dndy}         {\ensuremath{\mathrm{d}N_\mathrm{ch}/\mathrm{d}y}\xspace}
\newcommand{\dndeta}       {\ensuremath{\mathrm{d}N_\mathrm{ch}/\mathrm{d}\eta}\xspace}
\newcommand{\avdndeta}     {\ensuremath{\langle\dndeta\rangle}\xspace}
\newcommand{\dNdy}         {\ensuremath{\mathrm{d}N_\mathrm{ch}/\mathrm{d}y}\xspace}
\newcommand{\Npart}        {\ensuremath{N_\mathrm{part}}\xspace}
\newcommand{\Ncoll}        {\ensuremath{N_\mathrm{coll}}\xspace}
\newcommand{\dEdx}         {\ensuremath{\textrm{d}E/\textrm{d}x}\xspace}
\newcommand{\RpPb}         {\ensuremath{R_{\rm pPb}}\xspace}

\newcommand{\nineH}        {$\sqrt{s}~=~0.9$~Te\kern-.1emV\xspace}
\newcommand{\seven}        {$\sqrt{s}~=~7$~Te\kern-.1emV\xspace}
\newcommand{\twoH}         {$\sqrt{s}~=~0.2$~Te\kern-.1emV\xspace}
\newcommand{\twosevensix}  {$\sqrt{s}~=~2.76$~Te\kern-.1emV\xspace}
\newcommand{\five}         {$\sqrt{s}~=~5.02$~Te\kern-.1emV\xspace}
\newcommand{\twosevensixnn}{$\sqrt{s_{\mathrm{NN}}}~=~2.76$~Te\kern-.1emV\xspace}
\newcommand{\fivenn}       {$\sqrt{s_{\mathrm{NN}}}~=~5.02$~Te\kern-.1emV\xspace}
\newcommand{\LT}           {L{\'e}vy-Tsallis\xspace}
\newcommand{\GeVc}         {Ge\kern-.1emV/$c$\xspace}
\newcommand{\MeVc}         {Me\kern-.1emV/$c$\xspace}
\newcommand{\TeV}          {Te\kern-.1emV\xspace}
\newcommand{\GeV}          {Ge\kern-.1emV\xspace}
\newcommand{\MeV}          {Me\kern-.1emV\xspace}
\newcommand{\GeVmass}      {Ge\kern-.2emV/$c^2$\xspace}
\newcommand{\MeVmass}      {Me\kern-.2emV/$c^2$\xspace}
\newcommand{\lumi}         {\ensuremath{\mathcal{L}}\xspace}

\newcommand{\ITS}          {\rm{ITS}\xspace}
\newcommand{\TOF}          {\rm{TOF}\xspace}
\newcommand{\ZDC}          {\rm{ZDC}\xspace}
\newcommand{\ZDCs}         {\rm{ZDCs}\xspace}
\newcommand{\ZNA}          {\rm{ZNA}\xspace}
\newcommand{\ZNC}          {\rm{ZNC}\xspace}
\newcommand{\SPD}          {\rm{SPD}\xspace}
\newcommand{\SDD}          {\rm{SDD}\xspace}
\newcommand{\SSD}          {\rm{SSD}\xspace}
\newcommand{\TPC}          {\rm{TPC}\xspace}
\newcommand{\TRD}          {\rm{TRD}\xspace}
\newcommand{\VZERO}        {\rm{V0}\xspace}
\newcommand{\VZEROA}       {\rm{V0A}\xspace}
\newcommand{\VZEROC}       {\rm{V0C}\xspace}
\newcommand{\Vdecay} 	   {\ensuremath{V^{0}}\xspace}

\newcommand{\ee}           {\ensuremath{e^{+}e^{-}}} 
\newcommand{\pip}          {\ensuremath{\pi^{+}}\xspace}
\newcommand{\pim}          {\ensuremath{\pi^{-}}\xspace}
\newcommand{\kap}          {\ensuremath{\rm{K}^{+}}\xspace}
\newcommand{\kam}          {\ensuremath{\rm{K}^{-}}\xspace}
\newcommand{\pbar}         {\ensuremath{\rm\overline{p}}\xspace}
\newcommand{\kzero}        {\ensuremath{{\rm K}^{0}_{\rm{S}}}\xspace}
\newcommand{\lmb}          {\ensuremath{\Lambda}\xspace}
\newcommand{\almb}         {\ensuremath{\overline{\Lambda}}\xspace}
\newcommand{\Om}           {\ensuremath{\Omega^-}\xspace}
\newcommand{\Mo}           {\ensuremath{\overline{\Omega}^+}\xspace}
\newcommand{\X}            {\ensuremath{\Xi^-}\xspace}
\newcommand{\Ix}           {\ensuremath{\overline{\Xi}^+}\xspace}
\newcommand{\Xis}          {\ensuremath{\Xi^{\pm}}\xspace}
\newcommand{\Oms}          {\ensuremath{\Omega^{\pm}}\xspace}
\newcommand{\degree}       {\ensuremath{^{\rm o}}\xspace}

\begin{titlepage}
\PHyear{2019}       
\PHnumber{253}      
\PHdate{28 October}  

\title{Global baryon number conservation encoded in net-proton fluctuations measured in Pb--Pb collisions at $\sqrt{s_{\mathrm {NN}}} = 2.76$~TeV}
\ShortTitle{Global baryon number conservation encoded in net-proton fluctuations}   

\Collaboration{ALICE Collaboration\thanks{See Appendix~\ref{app:collab} for the list of collaboration members}}
\ShortAuthor{ALICE Collaboration} 

\begin{abstract}
Experimental results are presented on event-by-event net-proton fluctuation measurements in Pb--Pb collisions at $\sqrt{s_{\mathrm {NN}}} = 2.76$~TeV, recorded by the ALICE detector at the CERN LHC. These measurements have as their ultimate goal an experimental test of  Lattice QCD (LQCD) predictions on second and higher order cumulants of net-baryon distributions to search for critical behavior near the QCD phase boundary.  Before confronting them with LQCD predictions, account has to be taken of correlations stemming from baryon number conservation as well as fluctuations of participating nucleons. Both effects influence the experimental measurements and are usually not considered in theoretical calculations.  For the first time, it is shown that event-by-event baryon number conservation leads to subtle long-range correlations  arising from very early interactions in the collisions.  
\end{abstract}
\end{titlepage}

\setcounter{page}{2} 



Phase transitions in strongly interacting matter can be addressed by investigating the response of the system to external perturbations via measurements of fluctuations of conserved charges such as baryon number or electric charge~\cite{Koch:2008ia, Adamczyk:2013dal}. At LHC energies there would be, for vanishing light quark masses (u and d quarks),  a temperature-driven  second order phase transition between a hadron gas and a quark--gluon plasma~\cite{Stephanov:2004wx}. For realistic quark masses this transition becomes a smooth cross over~\cite{LQCD1,LQCD2}. However, because of the small masses of the current quarks,  one can still probe critical phenomena at LHC energies (vanishing baryon chemical potential) as reported in~\cite{Redlich1}. Indeed, recent LQCD calculations \cite{LQCD1,LQCD2} exhibit a rather strong signal for the existence of a pseudo-critical chiral temperature of about 156 MeV.  Moreover, this pseudo-critical temperature is in agreement with the chemical freeze-out  temperature as extracted by the analysis of hadron multiplicities~\cite{HRG,Andronic:2018qqt} measured by the ALICE experiment. This implies that the strongly interacting matter created in central collisions of Pb nuclei at LHC energies freezes out very near the chiral phase transition line. Hence, the singularities arising from the second order phase transition can be captured by measuring fluctuations of conserved charges such as the net-baryon number. Evaluated within the framework of the Hadron Resonance Gas (HRG), net-baryon distributions  coincide with the Skellam distribution, which is the probability distribution of the difference of two random variables, each generated from statistically independent Poisson distributions~\cite{Redlich:2012xf, skellam}. In fact, at the pseudo-critical temperature of 156 MeV, similar to the HRG framework, LQCD also predicts a Skellam behavior for the second cumulants of net-baryons, while the fourth cumulants of net-baryons from LQCD are significantly below the corresponding Skellam baseline~\cite{Bazavov:2018mes,Bazavov:2017dus}. 

Conserved quantities of course fluctuate only in sub-regions of the available total phase space of the reaction. In statistical mechanics they are, hence, analyzed within the Grand Canonical Ensemble (GCE)~\cite{StatLandau} formulation, where  only the average number of net-baryons are conserved.  In order to compare theoretical calculations within the GCE, such as the HRG~\cite{HRG} and LQCD~\cite{LQCD1,LQCD2}, with experimental results, the data are analyzed in different acceptance windows by imposing selection criteria on rapidity and/or  transverse momentum of the detected particles.  Indeed, if the selected acceptance window is too small, possible dynamical correlations will be washed out and the measured fluctuations will approach the Poisson limit~\cite{KochFluct}, implying that net-baryons will be distributed according to the Skellam distribution.

Recently the effects of limited acceptance  were studied~\cite{fluct_CE}. There, it was investigated under which conditions  net-baryon fluctuations depend on the size of the  acceptance. An obvious case is fluctuations caused by correlations due to baryon number conservation. To identify these and other long-range correlations it is interesting  to perform the experimental analysis as a function of the acceptance size. 

The analysis is set up by providing the necessary definitions. 
Given the number of baryons $(n_{\Brom})$ and antibaryons ($n_{\overline{\Brom}}$), the first and second cumulants of the net-baryon probability distribution $P(\Delta n_{\Brom})$, with $\Delta n_{\Brom} = n_{\Brom} - n_{\overline{\Brom}}$, are defined as
\eq{
\label{kumulants_definition}
  \kappa_{1}(\Delta n_{\Brom}) = \sum_{\Delta n_{\Brom}=-\infty}^{\infty}\Delta n_{\Brom}P(\Delta n_{\Brom})=\left<\Delta n_{\Brom}\right> , \\
  \kappa_{2} (\Delta n_{\Brom}) = \sum_{\Delta n_{\Brom}=-\infty}^{\infty}(\Delta n_{\Brom}-\left<\Delta n_{\Brom}\right>)^{2}P(\Delta n_{\Brom}) =  
  &\left<\left(\Delta n_{\Brom}-\left<\Delta n_{\Brom}\right>\right)^{2}\right>.}
 We note that, the first and second cumulants correspond to the expected value and the variance of net-baryon distribution, respectively. 
 The second cumulant can be represented  as a sum of the corresponding cumulants for
single baryons and antibaryons plus the correlation term for the joint probability distributions of baryons and antibaryons $P(n_{\Brom}, n_{\overline{\Brom}})$
\eq{
\label{sec_cum_2}
&\kappa_{2}(\Delta n_{\Brom})= \kappa_{2}(n_{\Brom})+\kappa_{2}(n_{\overline{\Brom}}) - 
2\left(\left<n_{\Brom}n_{\overline{\Brom}}\right> -
\left<n_{\Brom}\right>\left<n_{\overline{\Brom}}\right>\right),}
where the mixed moment $\left<n_{\Brom}n_{\overline{\Brom}}\right>$ is defined as
\eq{
\label{mixed_cumulant}
 \left<n_{\Brom}n_{\overline{\Brom}}\right>=\sum_{n_{\Brom}=0}^{\infty}\sum_{n_{\overline{\Brom}}=0}^{\infty}n_{\Brom}n_{\overline{\Brom}}P(n_{\Brom}, n_{\overline{\Brom}}). }

Equation~\ref{sec_cum_2} shows
that, for vanishing correlations between baryons and
antibaryons $(\left<n_{\Brom}n_{\overline{\Brom}}\right> = \left<n_{\Brom}\right>\left<n_{\overline{\Brom}}\right>)$, the second cumulant of  net-baryons  is exactly
equal to the sum of the corresponding second cumulants for baryons
and antibaryons.

The data presented below were obtained by analyzing about $13 \times 10^{6}$ minimum-bias (cf.~\cite{ALICE3} for definition) Pb--Pb events at  a center-of-mass energy per nucleon--nucleon pair of $\sqrt{s_{\mathrm {NN}}} = 2.76$~TeV recorded by the ALICE detector~\cite{ALICE} in the year 2010. Two forward scintillator arrays (V$0$) are located on either side of the interaction point and cover the pseudorapidity intervals $2.8 < \eta < 5.1$ and $-3.7 < \eta < -1.7$~\cite{Abelev:2013qoq}. A minimum-bias trigger condition is defined by requiring a combination of hits in the two innermost layers of the ITS and coincidence in both V$0$ detectors.  The event centrality is selected based  on the signal amplitudes in the V$0$ detectors~\cite{Abelev:2013qoq}.

The detectors used for track reconstruction are the Time Projection Chamber (TPC)~\cite{TPC} and the Inner Tracking System (ITS)~\cite{ITS}. In order to keep the tracking efficiency as high as possible, only the TPC detector was used for particle identification, while precise vertex determination was performed with the ITS detector. The track selection criteria used are described in Section 3.1 of \cite{Acharya:2017cpf}. Charged particle tracks with at least 80 out of maximum of 159 specific energy loss (d$E$/d$x$) samples in the TPC were used in this analysis for the best particle identification.  Moreover, in order to suppress contributions of secondary particles from weak decays, the distance-of-closest-approach (DCA) of the extrapolated track to the primary collision vertex was taken to be less than 2 cm along the beam direction. In the transverse plane, a more restrictive and transverse momentum ($\pt$) dependent DCA cut of less than (0.018 $\rm cm$ + 0.035$\pt^{-1.01}$) with $\pt$ in GeV/$c$, was imposed~\cite{ALICE2}.

Since the energy loss of charged particles in the gas volume of the TPC depends explicitly on the particle momentum ($p$), the analysis was performed in  momentum space. Correspondingly, the particle identification (PID) procedure consists of correlating particle momentum with the specific energy loss d$E$/d$x$. This allows the event-by-event fluctuation analysis to be performed with high overall reconstruction efficiency of about 80$\%$ for protons, almost independent of the collision centrality.  The latter is important because small efficiencies induce Poisson fluctuations. To ensure the best possible  d$E$/d$x$ resolution, the phase space coverage was  restricted to 0.6 $< p <$ 1.5 GeV/$c$ and $|\eta| < 0.8$ for the present analysis. The  corresponding  $\pt$ range at  $|\eta| = 0.8$ is about 0.45 $< \pt <$ 1.12 GeV/$c$, which gradually approaches the used momentum range towards midrapidity.  Moreover, a differential analysis is provided as function of $\Delta\eta$ in the range $\Delta\eta$ = 0.2 to 1.6. 

The cumulants of net-protons were then reconstructed using the Identity Method (IM)~\cite{Iden1, Iden2, Iden3, Iden4, ARSLANDOK2019162622, Acharya:2017cpf}. This method is designed to deal efficiently with the overlapping d$E$/d$x$ distributions of protons, kaons, pions and electrons considered in the present analysis. Their specific probability distributions were obtained by unfolding the moments of the measured multiplicity distributions for each particle species.  The IM counts proxies of particle multiplicities $W_{j}$ event-by-event
\begin{equation}
\label{eidentity1}
W_{j}=\sum_{i=1}^{n}\frac{\rho_{j}(x_{i})}{\rho(x_{i})},\;\:\:\:\; \rho(x_{i})=\sum_{j}\rho_{j}(x_{i}),
\end{equation}
where $j$ stands for a particle type, $x_{i}$ denotes the measured values of d$E$/d$x$ for a given track $i$ and $\rho_{j}(x)$ is the inclusive d$E$/d$x$ distribution of particle type $j$ within a specified phase space bin. The summation in Eq.~\ref{eidentity1} runs over all selected $n$ tracks in the given event. The pure and mixed moments of the $W_{j}$ distributions were calculated by averaging over all events, leading to the moments of the true multiplicity distributions. 

The IM is based on input of moments of $W_{j}$ distributions and inclusive d$E$/d$x$ fits in bins  of momentum  and pseudorapidity. The d$E$/d$x$ distributions were fit with  generalized Gaussian functions, taking into account non-Gaussian components of the experimental d$E$/d$x$ distributions. The fits of inclusive distributions of d$E$/d$x$ were performed separately for positively and negatively charged particles in 20 MeV/$c$ momentum and 0.1 units of $\eta$ bins. 

\begin{figure}[htb]
\centering
 \includegraphics[width=0.7\linewidth,clip=true]{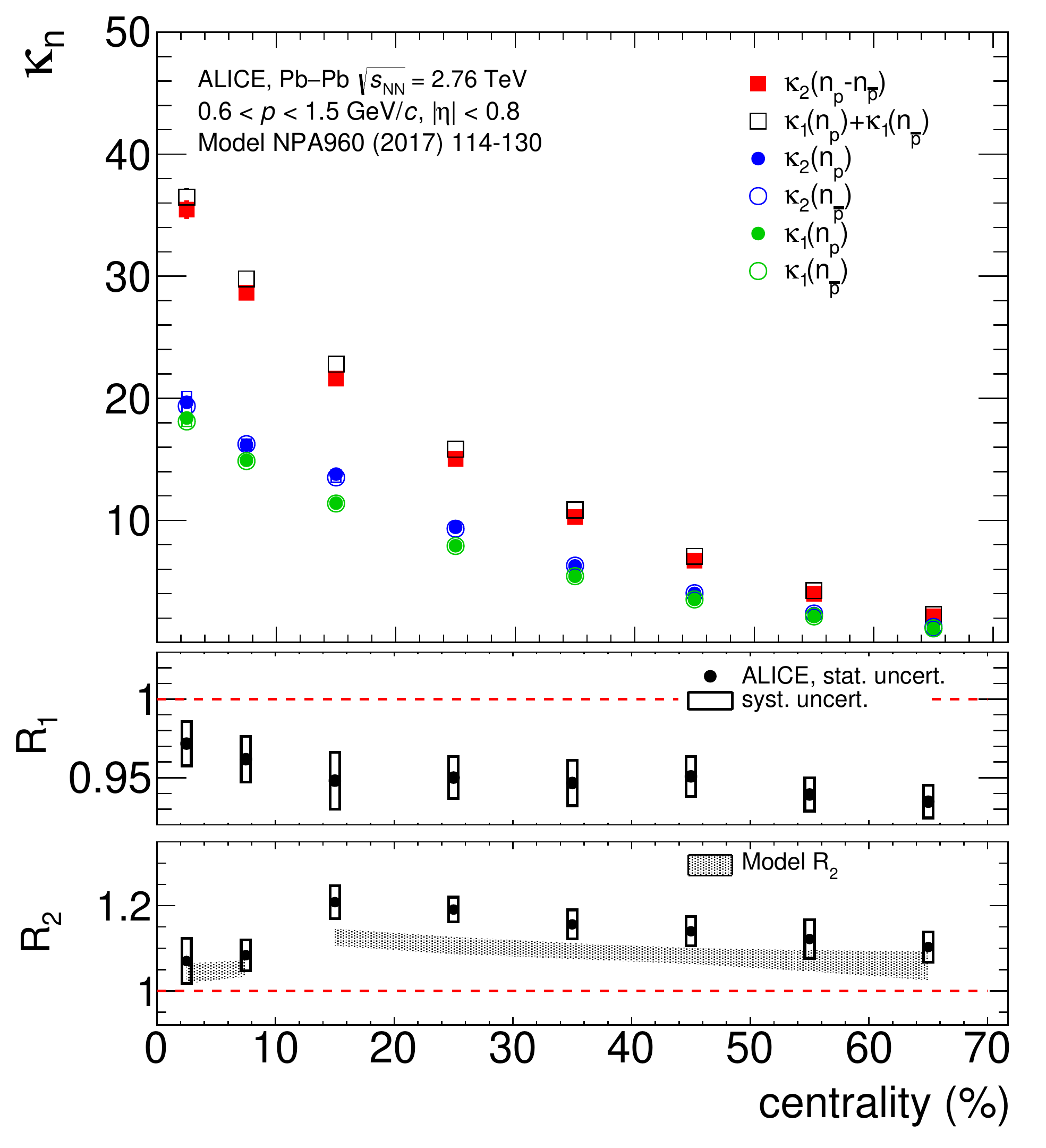}
 \caption{Measured second cumulants of net-proton distributions (red solid boxes) compared with the sum of the mean multiplicities (open squares). The second cumulants of single proton and antiproton distributions are presented with the filled and open  circles, respectively. The first cumulants of protons and antiprotons are hardly distinguishable because of the nearly equal mean numbers of protons and antiprotons at LHC energy.  In the middle and bottom  panels the normalized cumulants R$_{1}$and R$_{2}$ are presented. The band visible in the bottom panel is the prediction for R$_{2}$ in the presence of volume fluctuations~\cite{ourModel}.
 }
\label{fig_net_protons}
\end{figure}

In the upper panel of Fig.~\ref{fig_net_protons} the centrality dependence of the efficiency-corrected  second cumulants of net-protons  is compared with the sum of the mean multiplicities (first cumulants) of protons  and antiprotons. Also  included are the first and second cumulants of protons and antiprotons. The  efficiency correction for the cumulants is performed by using proton and anti-proton efficiencies in analytic formulas derived in~\cite{Bzdak:2012ab, Bzdak:2013pha} assuming binomial efficiency losses. The characteristics of the ALICE detector response and applied analysis procedure ensures that this assumption is fulfilled. The accuracy of the correction procedure was estimated to be on the percent level and is included in the systematic uncertainties. We note that possible corrections for volume fluctuations such as discussed in ~\cite{Sugiura:2019toh,Luo:2017faz} were not applied to the data since, at LHC energies, second cumulants of net protons, our main observable, are free from such effects~\cite{ourModel}.

The  subsample approach was chosen to estimate the statistical uncertainties of the reconstructed cumulants~\cite{Acharya:2017cpf, Anticic:2013htn}. In order to evaluate systematic uncertainties stemming from track selection criteria, the applied selection ranges were varied  around their nominal values. Other sources of systematic uncertainties originate from the parameters of the $\rho_{j}(x_{i})$ functions, entering Eq.~\ref{eidentity1}. The latter was estimated by hypothesis testing using the  Kolmogorov-Smirnov (K-S) statistics. For this purpose the parameters of the $\rho_{j}(x_{i})$ functions were varied by testing the null hypothesis that measured d$E$/d$x$ distributions and fit functions are similar within a given significance level of 10\% (cf.~\cite{Acharya:2017cpf, Anticic:2013htn} for details). The final systematic uncertainties on cumulants were computed by adding in quadrature the  maximum systematic variations from individual contributions.

By their definition, cumulants are extensive quantities, i.e., are proportional to the system volume. This also explains the centrality dependence of all cumulants, presented in the upper panel of Fig.~\ref{fig_net_protons}. To remove the system size dependence,  normalized cumulants R$_{1}$ and R$_{2}$ are introduced as

\begin{equation}
\mathrm{R}_{1}= \kappa_{2}(n_{\Prom} - n_{\overline{\Prom}})/< n_{\Prom} + n_{\overline{\Prom}} >,\,\,\ \mathrm{R}_{2}= \kappa_{2}(n_{\Prom}) /< n_{\Prom} >.
\label{normCum}
\end{equation}

In the middle and bottom panels of Fig.~\ref{fig_net_protons}, deviations from unity are visible for both R$_{1}$ and R$_{2}$. Moreover, the amount of deviation for R$_{2}$ is about twice as large compared to that of R$_{1}$. 
 
In order to shed light on these observations, the results are compared with predictions from a model constructed recently~\cite{ourModel}, in which participant fluctuations are included following the analysis of the ALICE centrality selection~\cite{Abelev:2013qoq}. Within  uncertainties, the model predictions are fully consistent with the measured R$_{2}$ values, lending support to the interpretation that volume fluctuations are at the origin of the observed deviation. This is also supported by the observation of a small structure observed in the  10--20\% centrality class, where, compared to the first two centrality classes, the centrality bin width is doubled.

On the other hand, by construction, for vanishing net-proton numbers, R$_{1}$ should not contain any contributions from volume fluctuations, i.e., the values of R$_{1}$ obtained from the model should be consistent with unity~\cite{ourModel}. In fact at LHC energies R$_{1}$ becomes a strongly intensive quantity~\cite{Gorenstein:2011vq}.
The origin for the deviation of the measured R$_{1}$ values from unity must therefore be beyond the volume fluctuations scenario.  To further understand these differences, the acceptance dependence is studied.

 \begin{figure}[htb]
\centering
 \includegraphics[width=0.7\linewidth,clip=true]{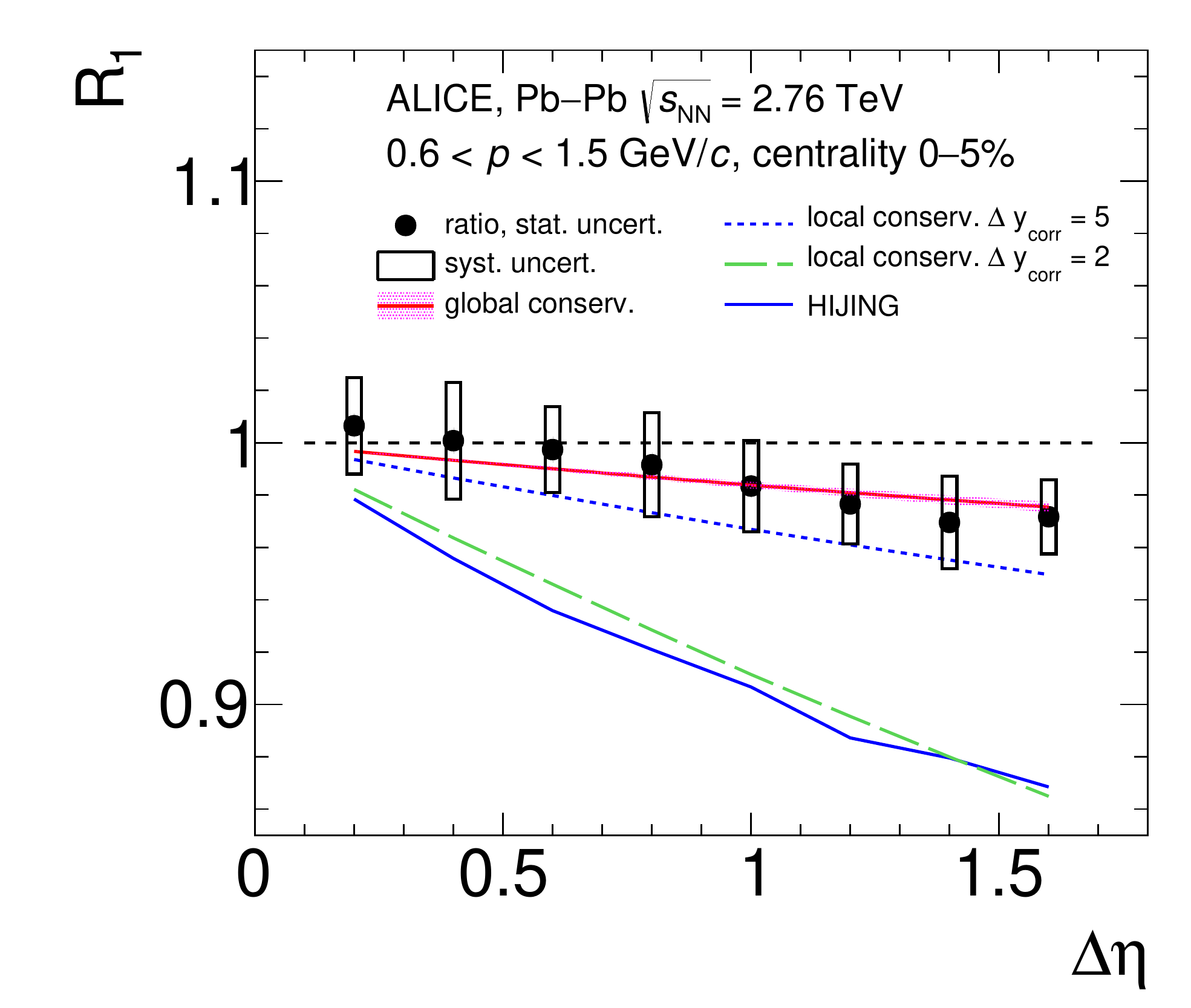} 
 \caption{Pseudorapidity  dependence of the normalized second cumulants of net-protons R$_1$. Global baryon number conservation is depicted as the pink band. The dashed lines represent the predictions from the model with local baryon number conservation~\cite{Braun-Munzinger:2019yxj}. The blue solid line, represents the prediction using the HIJING generator. }
\label{fig_acc_dep}
\end{figure}  

The analysis is performed in eight different pseudorapidity intervals from
$|\eta| < 0.1$ up to $|\eta| < 0.8$ in steps of 0.1. The obtained
normalized second cumulants R$_1$ of net-protons are presented in Fig.~\ref{fig_acc_dep}. Again the data are below unity, with the deviation linearly increasing with increasing acceptance.

Such a behavior was predicted based on the assumption of global baryon number conservation~\cite{Mrowczynski:2001mm, Bzdak:2012an, ourModel}, which  induces correlations between protons and antiprotons leading to the following dependence on the acceptance factor $\alpha$ 

\begin{equation}
\mathrm{R}_{1}=1-\alpha,
\label{conserv2}
\end{equation}

where $\alpha = \left< n_{\Prom}\right> / \left< N_{\Brom}^{4\pi}\right>$
with $\left<n_{\Prom}\right>$ and $\left< N_{\Brom}^{4\pi}\right>$ referring to the mean number of protons inside the acceptance and the mean number of baryons in full phase space. It should be further noted that, for non-central collisions, baryon transport to mid-rapidity has to be taken into account, which is rather model dependent. In order to avoid the model dependence, the comparison is performed only for the central events and in the estimate of the alpha parameter only produced baryons are used. 
In doing so, the  number of baryons are used in the pseudorapidity range of $|\eta| < 0.5$ as reported in ~\cite{ALICE3, Abelev:2013xaa, ABELEV:2013zaa, fluct_exp}.  Next, using HIJING and AMPT simulations, estimates were obtained  for the total average number of baryons in
full phase space. The average number of protons $\left< n_{\Prom}\right>$ entering into the definition of $\alpha$ (cf. Eq.~\ref{conserv2}) was taken from the current analysis for each
pseudorapidity interval. Finally, using these values of $\alpha$, the pink
band in Fig.~\ref{fig_acc_dep} is calculated with Eq.~\ref{conserv2}. The finite width of the band reflects the difference between predictions of the two event generators.

Inspection of Fig.~\ref{fig_acc_dep} shows that, for small pseudorapidity ranges of $|\eta| <0.4$ corresponding to $\Delta\eta<0.8$, the experimentally measured net-proton distributions closely follow a Skellam distribution. This  agreement is expected  because of the small acceptance window as discussed above. For $\Delta \eta > 0.8$,  
deviations from the Skellam distribution are observed. The amount of deviation is small but significant and in good agreement with the prediction assuming global baryon number conservation. The observed deviation is therefore consistent with the assumption of global baryon number conservation, i.e. conservation within the full phase space. 

On the other hand, local baryon number conservation may induce additional correlations between protons and antiprotons, which would lead to a further reduction of the measured $\kappa_{2}(n_{\Prom} - n_{\overline{\Prom}})$~\cite{Braun-Munzinger:2019yxj}.
In Fig.~\ref{fig_acc_dep} the data are compared to the predictions from an analysis of effects of local baryon number conservation for different values of correlation width $\Delta y_{\mathrm{corr}}$ between protons and antiprotons.Within the experimental uncertainties the data are best described with the assumption of  global baryon number conservation, which corresponds to the correlation width $\Delta y_{\mathrm{corr}} = 2|y_{\mathrm{beam}}|$ but, within one standard deviation (1.56 for the last point at $\Delta \eta = 1.6$), the data are also consistent with a large correlation width of $\Delta y_{\mathrm{corr}} = 5$~\cite{Braun-Munzinger:2019yxj}. We find that for $\Delta y_{\mathrm{corr}} = 4.5$, with a 5$\%$ significance level, the last point is not consistent with the experimental data.  The results from the HIJING event generator (cf. blue solid line in Fig.~\ref{fig_acc_dep}), which can be described with $\Delta y_{\mathrm{corr}} = 2$, and from a recent study reported in~\cite{Pruneau:2019baa}  would imply much stronger correlations between protons and antiprotons, not consistent with the experimental data. We note here that correlations arising from baryon or charge conservation have also been analyzed in the framework of balance functions~\cite{Pratt:2003ij,Abelev:2013csa}. Such an analysis could also shed interesting light on global vs. local baryon conservation.

From the present results it is concluded that effects due to local baryon number conservation are not large, if present at all in second cumulants of net-protons. The large correlation length observed in the data implies that the normalized second cumulant R1 is determined by collisions in the very early phase of the Pb--Pb interaction~\cite{Dumitru:2008wn}. We note that long range rapidity correlations were investigated in other contexts in~\cite{Capella:1978rg, Vechernin:2012bz}. The search for critical behavior, as predicted for higher cumulants of net-baryon distributions~\cite{Bazavov:2017dus, Almasi:2017bhq}, will be the topic of future investigations.


\newenvironment{acknowledgement}{\relax}{\relax}
\begin{acknowledgement}
\section*{Acknowledgements}

The ALICE Collaboration would like to thank all its engineers and technicians for their invaluable contributions to the construction of the experiment and the CERN accelerator teams for the outstanding performance of the LHC complex.
The ALICE Collaboration gratefully acknowledges the resources and support provided by all Grid centres and the Worldwide LHC Computing Grid (WLCG) collaboration.
The ALICE Collaboration acknowledges the following funding agencies for their support in building and running the ALICE detector:
A. I. Alikhanyan National Science Laboratory (Yerevan Physics Institute) Foundation (ANSL), State Committee of Science and World Federation of Scientists (WFS), Armenia;
Austrian Academy of Sciences, Austrian Science Fund (FWF): [M 2467-N36] and Nationalstiftung f\"{u}r Forschung, Technologie und Entwicklung, Austria;
Ministry of Communications and High Technologies, National Nuclear Research Center, Azerbaijan;
Conselho Nacional de Desenvolvimento Cient\'{\i}fico e Tecnol\'{o}gico (CNPq), Financiadora de Estudos e Projetos (Finep), Funda\c{c}\~{a}o de Amparo \`{a} Pesquisa do Estado de S\~{a}o Paulo (FAPESP) and Universidade Federal do Rio Grande do Sul (UFRGS), Brazil;
Ministry of Education of China (MOEC) , Ministry of Science \& Technology of China (MSTC) and National Natural Science Foundation of China (NSFC), China;
Ministry of Science and Education and Croatian Science Foundation, Croatia;
Centro de Aplicaciones Tecnol\'{o}gicas y Desarrollo Nuclear (CEADEN), Cubaenerg\'{\i}a, Cuba;
Ministry of Education, Youth and Sports of the Czech Republic, Czech Republic;
The Danish Council for Independent Research | Natural Sciences, the VILLUM FONDEN and Danish National Research Foundation (DNRF), Denmark;
Helsinki Institute of Physics (HIP), Finland;
Commissariat \`{a} l'Energie Atomique (CEA), Institut National de Physique Nucl\'{e}aire et de Physique des Particules (IN2P3) and Centre National de la Recherche Scientifique (CNRS) and R\'{e}gion des  Pays de la Loire, France;
Bundesministerium f\"{u}r Bildung und Forschung (BMBF) and GSI Helmholtzzentrum f\"{u}r Schwerionenforschung GmbH, Germany;
General Secretariat for Research and Technology, Ministry of Education, Research and Religions, Greece;
National Research, Development and Innovation Office, Hungary;
Department of Atomic Energy Government of India (DAE), Department of Science and Technology, Government of India (DST), University Grants Commission, Government of India (UGC) and Council of Scientific and Industrial Research (CSIR), India;
Indonesian Institute of Science, Indonesia;
Centro Fermi - Museo Storico della Fisica e Centro Studi e Ricerche Enrico Fermi and Istituto Nazionale di Fisica Nucleare (INFN), Italy;
Institute for Innovative Science and Technology , Nagasaki Institute of Applied Science (IIST), Japanese Ministry of Education, Culture, Sports, Science and Technology (MEXT) and Japan Society for the Promotion of Science (JSPS) KAKENHI, Japan;
Consejo Nacional de Ciencia (CONACYT) y Tecnolog\'{i}a, through Fondo de Cooperaci\'{o}n Internacional en Ciencia y Tecnolog\'{i}a (FONCICYT) and Direcci\'{o}n General de Asuntos del Personal Academico (DGAPA), Mexico;
Nederlandse Organisatie voor Wetenschappelijk Onderzoek (NWO), Netherlands;
The Research Council of Norway, Norway;
Commission on Science and Technology for Sustainable Development in the South (COMSATS), Pakistan;
Pontificia Universidad Cat\'{o}lica del Per\'{u}, Peru;
Ministry of Science and Higher Education and National Science Centre, Poland;
Korea Institute of Science and Technology Information and National Research Foundation of Korea (NRF), Republic of Korea;
Ministry of Education and Scientific Research, Institute of Atomic Physics and Ministry of Research and Innovation and Institute of Atomic Physics, Romania;
Joint Institute for Nuclear Research (JINR), Ministry of Education and Science of the Russian Federation, National Research Centre Kurchatov Institute, Russian Science Foundation and Russian Foundation for Basic Research, Russia;
Ministry of Education, Science, Research and Sport of the Slovak Republic, Slovakia;
National Research Foundation of South Africa, South Africa;
Swedish Research Council (VR) and Knut \& Alice Wallenberg Foundation (KAW), Sweden;
European Organization for Nuclear Research, Switzerland;
Suranaree University of Technology (SUT), National Science and Technology Development Agency (NSDTA) and Office of the Higher Education Commission under NRU project of Thailand, Thailand;
Turkish Atomic Energy Agency (TAEK), Turkey;
National Academy of  Sciences of Ukraine, Ukraine;
Science and Technology Facilities Council (STFC), United Kingdom;
National Science Foundation of the United States of America (NSF) and United States Department of Energy, Office of Nuclear Physics (DOE NP), United States of America.
\end{acknowledgement}

\bibliographystyle{utphys}   
\bibliography{bibliography}

\newpage
\appendix

%
%

\section{The ALICE Collaboration}
\label{app:collab}

\begingroup
\small
\begin{flushleft}
S.~Acharya\Irefn{org141}\And 
D.~Adamov\'{a}\Irefn{org94}\And 
A.~Adler\Irefn{org74}\And 
J.~Adolfsson\Irefn{org80}\And 
M.M.~Aggarwal\Irefn{org99}\And 
G.~Aglieri Rinella\Irefn{org33}\And 
M.~Agnello\Irefn{org30}\And 
N.~Agrawal\Irefn{org10}\textsuperscript{,}\Irefn{org53}\And 
Z.~Ahammed\Irefn{org141}\And 
S.~Ahmad\Irefn{org16}
% ...
\And
\renewcommand\labelenumi{\textsuperscript{\theenumi}~}

\section*{Affiliation notes}
\renewcommand\theenumi{\roman{enumi}}
\begin{Authlist}
\item \Adef{org*}Deceased
\item \Adef{orgI}Dipartimento DET del Politecnico di Torino, Turin, Italy
\item \Adef{orgII}M.V. Lomonosov Moscow State University, D.V. Skobeltsyn Institute of Nuclear, Physics, Moscow, Russia
\item \Adef{orgIII}Department of Applied Physics, Aligarh Muslim University, Aligarh, India
\item \Adef{orgIV}Institute of Theoretical Physics, University of Wroclaw, Poland
\end{Authlist}

\section*{Collaboration Institutes}
\renewcommand\theenumi{\arabic{enumi}~}
\begin{Authlist}
\item \Idef{org1}A.I. Alikhanyan National Science Laboratory (Yerevan Physics Institute) Foundation, Yerevan, Armenia
\item \Idef{org2}Bogolyubov Institute for Theoretical Physics, National Academy of Sciences of Ukraine, Kiev, Ukraine
\item \Idef{org3}Bose Institute, Department of Physics  and Centre for Astroparticle Physics and Space Science (CAPSS), Kolkata, India
\item \Idef{org4}Budker Institute for Nuclear Physics, Novosibirsk, Russia
\item \Idef{org5}California Polytechnic State University, San Luis Obispo, California, United States
\item \Idef{org6}Central China Normal University, Wuhan, China
\item \Idef{org7}Centre de Calcul de l'IN2P3, Villeurbanne, Lyon, France
\item \Idef{org8}Centro de Aplicaciones Tecnol\'{o}gicas y Desarrollo Nuclear (CEADEN), Havana, Cuba
\item \Idef{org9}Centro de Investigaci\'{o}n y de Estudios Avanzados (CINVESTAV), Mexico City and M\'{e}rida, Mexico
\item \Idef{org10}Centro Fermi - Museo Storico della Fisica e Centro Studi e Ricerche ``Enrico Fermi', Rome, Italy
\item \Idef{org11}Chicago State University, Chicago, Illinois, United States
\item \Idef{org12}China Institute of Atomic Energy, Beijing, China
\item \Idef{org13}Comenius University Bratislava, Faculty of Mathematics, Physics and Informatics, Bratislava, Slovakia
\item \Idef{org14}COMSATS University Islamabad, Islamabad, Pakistan
\item \Idef{org15}Creighton University, Omaha, Nebraska, United States
\item \Idef{org16}Department of Physics, Aligarh Muslim University, Aligarh, India
\item \Idef{org17}Department of Physics, Pusan National University, Pusan, Republic of Korea
\item \Idef{org18}Department of Physics, Sejong University, Seoul, Republic of Korea
\item \Idef{org19}Department of Physics, University of California, Berkeley, California, United States
\item \Idef{org20}Department of Physics, University of Oslo, Oslo, Norway
\item \Idef{org21}Department of Physics and Technology, University of Bergen, Bergen, Norway
\item \Idef{org22}Dipartimento di Fisica dell'Universit\`{a} 'La Sapienza' and Sezione INFN, Rome, Italy
\item \Idef{org23}Dipartimento di Fisica dell'Universit\`{a} and Sezione INFN, Cagliari, Italy
\item \Idef{org24}Dipartimento di Fisica dell'Universit\`{a} and Sezione INFN, Trieste, Italy
\item \Idef{org25}Dipartimento di Fisica dell'Universit\`{a} and Sezione INFN, Turin, Italy
\item \Idef{org26}Dipartimento di Fisica e Astronomia dell'Universit\`{a} and Sezione INFN, Bologna, Italy
\item \Idef{org27}Dipartimento di Fisica e Astronomia dell'Universit\`{a} and Sezione INFN, Catania, Italy
\item \Idef{org28}Dipartimento di Fisica e Astronomia dell'Universit\`{a} and Sezione INFN, Padova, Italy
\item \Idef{org29}Dipartimento di Fisica `E.R.~Caianiello' dell'Universit\`{a} and Gruppo Collegato INFN, Salerno, Italy
\item \Idef{org30}Dipartimento DISAT del Politecnico and Sezione INFN, Turin, Italy
\item \Idef{org31}Dipartimento di Scienze e Innovazione Tecnologica dell'Universit\`{a} del Piemonte Orientale and INFN Sezione di Torino, Alessandria, Italy
\item \Idef{org32}Dipartimento Interateneo di Fisica `M.~Merlin' and Sezione INFN, Bari, Italy
\item \Idef{org33}European Organization for Nuclear Research (CERN), Geneva, Switzerland
\item \Idef{org34}Faculty of Electrical Engineering, Mechanical Engineering and Naval Architecture, University of Split, Split, Croatia
\item \Idef{org35}Faculty of Engineering and Science, Western Norway University of Applied Sciences, Bergen, Norway
\item \Idef{org36}Faculty of Nuclear Sciences and Physical Engineering, Czech Technical University in Prague, Prague, Czech Republic
\item \Idef{org37}Faculty of Science, P.J.~\v{S}af\'{a}rik University, Ko\v{s}ice, Slovakia
\item \Idef{org38}Frankfurt Institute for Advanced Studies, Johann Wolfgang Goethe-Universit\"{a}t Frankfurt, Frankfurt, Germany
\item \Idef{org39}Fudan University, Shanghai, China
\item \Idef{org40}Gangneung-Wonju National University, Gangneung, Republic of Korea
\item \Idef{org41}Gauhati University, Department of Physics, Guwahati, India
\item \Idef{org42}Helmholtz-Institut f\"{u}r Strahlen- und Kernphysik, Rheinische Friedrich-Wilhelms-Universit\"{a}t Bonn, Bonn, Germany
\item \Idef{org43}Helsinki Institute of Physics (HIP), Helsinki, Finland
\item \Idef{org44}High Energy Physics Group,  Universidad Aut\'{o}noma de Puebla, Puebla, Mexico
\item \Idef{org45}Hiroshima University, Hiroshima, Japan
\item \Idef{org46}Hochschule Worms, Zentrum  f\"{u}r Technologietransfer und Telekommunikation (ZTT), Worms, Germany
\item \Idef{org47}Horia Hulubei National Institute of Physics and Nuclear Engineering, Bucharest, Romania
\item \Idef{org48}Indian Institute of Technology Bombay (IIT), Mumbai, India
\item \Idef{org49}Indian Institute of Technology Indore, Indore, India
\item \Idef{org50}Indonesian Institute of Sciences, Jakarta, Indonesia
\item \Idef{org51}INFN, Laboratori Nazionali di Frascati, Frascati, Italy
\item \Idef{org52}INFN, Sezione di Bari, Bari, Italy
\item \Idef{org53}INFN, Sezione di Bologna, Bologna, Italy
\item \Idef{org54}INFN, Sezione di Cagliari, Cagliari, Italy
\item \Idef{org55}INFN, Sezione di Catania, Catania, Italy
\item \Idef{org56}INFN, Sezione di Padova, Padova, Italy
\item \Idef{org57}INFN, Sezione di Roma, Rome, Italy
\item \Idef{org58}INFN, Sezione di Torino, Turin, Italy
\item \Idef{org59}INFN, Sezione di Trieste, Trieste, Italy
\item \Idef{org60}Inha University, Incheon, Republic of Korea
\item \Idef{org61}Institut de Physique Nucl\'{e}aire d'Orsay (IPNO), Institut National de Physique Nucl\'{e}aire et de Physique des Particules (IN2P3/CNRS), Universit\'{e} de Paris-Sud, Universit\'{e} Paris-Saclay, Orsay, France
\item \Idef{org62}Institute for Nuclear Research, Academy of Sciences, Moscow, Russia
\item \Idef{org63}Institute for Subatomic Physics, Utrecht University/Nikhef, Utrecht, Netherlands
\item \Idef{org64}Institute of Experimental Physics, Slovak Academy of Sciences, Ko\v{s}ice, Slovakia
\item \Idef{org65}Institute of Physics, Homi Bhabha National Institute, Bhubaneswar, India
\item \Idef{org66}Institute of Physics of the Czech Academy of Sciences, Prague, Czech Republic
\item \Idef{org67}Institute of Space Science (ISS), Bucharest, Romania
\item \Idef{org68}Institut f\"{u}r Kernphysik, Johann Wolfgang Goethe-Universit\"{a}t Frankfurt, Frankfurt, Germany
\item \Idef{org69}Instituto de Ciencias Nucleares, Universidad Nacional Aut\'{o}noma de M\'{e}xico, Mexico City, Mexico
\item \Idef{org70}Instituto de F\'{i}sica, Universidade Federal do Rio Grande do Sul (UFRGS), Porto Alegre, Brazil
\item \Idef{org71}Instituto de F\'{\i}sica, Universidad Nacional Aut\'{o}noma de M\'{e}xico, Mexico City, Mexico
\item \Idef{org72}iThemba LABS, National Research Foundation, Somerset West, South Africa
\item \Idef{org73}Jeonbuk National University, Jeonju, Republic of Korea
\item \Idef{org74}Johann-Wolfgang-Goethe Universit\"{a}t Frankfurt Institut f\"{u}r Informatik, Fachbereich Informatik und Mathematik, Frankfurt, Germany
\item \Idef{org75}Joint Institute for Nuclear Research (JINR), Dubna, Russia
\item \Idef{org76}Korea Institute of Science and Technology Information, Daejeon, Republic of Korea
\item \Idef{org77}KTO Karatay University, Konya, Turkey
\item \Idef{org78}Laboratoire de Physique Subatomique et de Cosmologie, Universit\'{e} Grenoble-Alpes, CNRS-IN2P3, Grenoble, France
\item \Idef{org79}Lawrence Berkeley National Laboratory, Berkeley, California, United States
\item \Idef{org80}Lund University Department of Physics, Division of Particle Physics, Lund, Sweden
\item \Idef{org81}Nagasaki Institute of Applied Science, Nagasaki, Japan
\item \Idef{org82}Nara Women{'}s University (NWU), Nara, Japan
\item \Idef{org83}National and Kapodistrian University of Athens, School of Science, Department of Physics , Athens, Greece
\item \Idef{org84}National Centre for Nuclear Research, Warsaw, Poland
\item \Idef{org85}National Institute of Science Education and Research, Homi Bhabha National Institute, Jatni, India
\item \Idef{org86}National Nuclear Research Center, Baku, Azerbaijan
\item \Idef{org87}National Research Centre Kurchatov Institute, Moscow, Russia
\item \Idef{org88}Niels Bohr Institute, University of Copenhagen, Copenhagen, Denmark
\item \Idef{org89}Nikhef, National institute for subatomic physics, Amsterdam, Netherlands
\item \Idef{org90}NRC Kurchatov Institute IHEP, Protvino, Russia
\item \Idef{org91}NRC «Kurchatov Institute»  - ITEP, Moscow, Russia
\item \Idef{org92}NRNU Moscow Engineering Physics Institute, Moscow, Russia
\item \Idef{org93}Nuclear Physics Group, STFC Daresbury Laboratory, Daresbury, United Kingdom
\item \Idef{org94}Nuclear Physics Institute of the Czech Academy of Sciences, \v{R}e\v{z} u Prahy, Czech Republic
\item \Idef{org95}Oak Ridge National Laboratory, Oak Ridge, Tennessee, United States
\item \Idef{org96}Ohio State University, Columbus, Ohio, United States
\item \Idef{org97}Petersburg Nuclear Physics Institute, Gatchina, Russia
\item \Idef{org98}Physics department, Faculty of science, University of Zagreb, Zagreb, Croatia
\item \Idef{org99}Physics Department, Panjab University, Chandigarh, India
\item \Idef{org100}Physics Department, University of Jammu, Jammu, India
\item \Idef{org101}Physics Department, University of Rajasthan, Jaipur, India
\item \Idef{org102}Physikalisches Institut, Eberhard-Karls-Universit\"{a}t T\"{u}bingen, T\"{u}bingen, Germany
\item \Idef{org103}Physikalisches Institut, Ruprecht-Karls-Universit\"{a}t Heidelberg, Heidelberg, Germany
\item \Idef{org104}Physik Department, Technische Universit\"{a}t M\"{u}nchen, Munich, Germany
\item \Idef{org105}Politecnico di Bari, Bari, Italy
\item \Idef{org106}Research Division and ExtreMe Matter Institute EMMI, GSI Helmholtzzentrum f\"ur Schwerionenforschung GmbH, Darmstadt, Germany
\item \Idef{org107}Rudjer Bo\v{s}kovi\'{c} Institute, Zagreb, Croatia
\item \Idef{org108}Russian Federal Nuclear Center (VNIIEF), Sarov, Russia
\item \Idef{org109}Saha Institute of Nuclear Physics, Homi Bhabha National Institute, Kolkata, India
\item \Idef{org110}School of Physics and Astronomy, University of Birmingham, Birmingham, United Kingdom
\item \Idef{org111}Secci\'{o}n F\'{\i}sica, Departamento de Ciencias, Pontificia Universidad Cat\'{o}lica del Per\'{u}, Lima, Peru
\item \Idef{org112}St. Petersburg State University, St. Petersburg, Russia
\item \Idef{org113}Stefan Meyer Institut f\"{u}r Subatomare Physik (SMI), Vienna, Austria
\item \Idef{org114}SUBATECH, IMT Atlantique, Universit\'{e} de Nantes, CNRS-IN2P3, Nantes, France
\item \Idef{org115}Suranaree University of Technology, Nakhon Ratchasima, Thailand
\item \Idef{org116}Technical University of Ko\v{s}ice, Ko\v{s}ice, Slovakia
\item \Idef{org117}Technische Universit\"{a}t M\"{u}nchen, Excellence Cluster 'Universe', Munich, Germany
\item \Idef{org118}The Henryk Niewodniczanski Institute of Nuclear Physics, Polish Academy of Sciences, Cracow, Poland
\item \Idef{org119}The University of Texas at Austin, Austin, Texas, United States
\item \Idef{org120}Universidad Aut\'{o}noma de Sinaloa, Culiac\'{a}n, Mexico
\item \Idef{org121}Universidade de S\~{a}o Paulo (USP), S\~{a}o Paulo, Brazil
\item \Idef{org122}Universidade Estadual de Campinas (UNICAMP), Campinas, Brazil
\item \Idef{org123}Universidade Federal do ABC, Santo Andre, Brazil
\item \Idef{org124}University of Cape Town, Cape Town, South Africa
\item \Idef{org125}University of Houston, Houston, Texas, United States
\item \Idef{org126}University of Jyv\"{a}skyl\"{a}, Jyv\"{a}skyl\"{a}, Finland
\item \Idef{org127}University of Liverpool, Liverpool, United Kingdom
\item \Idef{org128}University of Science and Techonology of China, Hefei, China
\item \Idef{org129}University of South-Eastern Norway, Tonsberg, Norway
\item \Idef{org130}University of Tennessee, Knoxville, Tennessee, United States
\item \Idef{org131}University of the Witwatersrand, Johannesburg, South Africa
\item \Idef{org132}University of Tokyo, Tokyo, Japan
\item \Idef{org133}University of Tsukuba, Tsukuba, Japan
\item \Idef{org134}Universit\'{e} Clermont Auvergne, CNRS/IN2P3, LPC, Clermont-Ferrand, France
\item \Idef{org135}Universit\'{e} de Lyon, Universit\'{e} Lyon 1, CNRS/IN2P3, IPN-Lyon, Villeurbanne, Lyon, France
\item \Idef{org136}Universit\'{e} de Strasbourg, CNRS, IPHC UMR 7178, F-67000 Strasbourg, France, Strasbourg, France
\item \Idef{org137}Universit\'{e} Paris-Saclay Centre d'Etudes de Saclay (CEA), IRFU, D\'{e}partment de Physique Nucl\'{e}aire (DPhN), Saclay, France
\item \Idef{org138}Universit\`{a} degli Studi di Foggia, Foggia, Italy
\item \Idef{org139}Universit\`{a} degli Studi di Pavia, Pavia, Italy
\item \Idef{org140}Universit\`{a} di Brescia, Brescia, Italy
\item \Idef{org141}Variable Energy Cyclotron Centre, Homi Bhabha National Institute, Kolkata, India
\item \Idef{org142}Warsaw University of Technology, Warsaw, Poland
\item \Idef{org143}Wayne State University, Detroit, Michigan, United States
\item \Idef{org144}Westf\"{a}lische Wilhelms-Universit\"{a}t M\"{u}nster, Institut f\"{u}r Kernphysik, M\"{u}nster, Germany
\item \Idef{org145}Wigner Research Centre for Physics, Budapest, Hungary
\item \Idef{org146}Yale University, New Haven, Connecticut, United States
\item \Idef{org147}Yonsei University, Seoul, Republic of Korea
\end{Authlist}
\endgroup
  
\end{document}